\documentclass[10pt]{article}
\usepackage[paperwidth=210mm,paperheight=297mm,top=28mm,bottom=22mm,left=25mm,right=25mm,dvips]{geometry}
\usepackage{stmaryrd}
\usepackage{bbm}
\usepackage{mathrsfs}
\usepackage{amsfonts}
\usepackage{amssymb}
\usepackage{amsmath}
\usepackage{amsthm}
\usepackage{cases}
\usepackage{indentfirst}
\usepackage{lscape}
\usepackage{graphicx}
\usepackage{chngcntr}

\allowdisplaybreaks[4]
\begin{document}

\title{ {\bf Symmetry analysis for time-fractional convection-diffusion equation}
\footnotetext{*Email: mathzj@zjut.edu.cn}
\author{ Junjun Zhang$^{a}$, Jun Zhang$^{a,b,*}$}}
\date{}
\maketitle
\begin{flushleft}
{\small
$^a$ Department of Applied Mathematics, Zhejiang
University of Technology, Hangzhou 310023, China\\
$^b$ Centre for Mathematics and Its Applications, The Australian National University, Canberra, ACT 0200, Australia}
\end{flushleft}

{\bf Abstract:} The time-fractional convection-diffusion equation is performed by Lie symmetry analysis method which involves the Riemann-Liouville time-fractional derivative of the order $\alpha\in(0,2)$. In eight cases, the symmetries are obtained and similarity reductions of the equation are deduced by means of symmetry. It is shown that the fractional equation can be reduced into fractional ordinary differential equations. Some group invariant solutions in explicit form are obtained in some cases.


\section{Introduction}
It is well known  that Lie symmetry theory plays a significant role in the analysis of  differential  equations[1-5]. The basic idea of this method is that the infinitesimal transformation leaves the set of solution manifold of the considered differential equation invariant. This efficient method invented by Sophus Lie is a highly algorithmic process, and it often involves lengthy symbolic computation. The method systematically unifies and extends well-known techniques to construct explicit solutions for differential equations, especially for nonlinear differential equations. In recent years this method has a successful extension to discrete systems exhibiting solitons governed by nonlinear partial differential-difference equations and pure difference equations[6-8].

Recently the study of fractional differential equations(FDEs), as generalizations of classical integer order differential equations, has attracted much attention due to an exact description of nonlinear phenomena in fluid mechanics, viscoelasticity, biology, physics, engineering and other areas of science[9-12]. However, unlike the classical integer order derivatives, there exists a number of different definitions of fractional order derivatives and corresponding FDEs. These definition differences lead to the FDEs having similar form but significantly different properties. It means that there exists no well-defined method to analyze them systematically. As a consequence, several different analytical methods such as differential transform method[13], Adomian decomposition method[14,15], invariant subspace method[16], Green function approach[17] and symmetry analysis[18-26] have been formulated to reduce and solve FDEs.

In this article, we consider the following time-fractional convection-diffusion equation
\begin{equation} \label{eq:1}
\frac{\partial^\alpha {u}}{\partial{t}^\alpha}=(D(u)u_{x})_{x}+P(u)u_{x},0<\alpha<2,
\end{equation}
where $\frac{\partial^\alpha{u}}{\partial{t}^\alpha}$ is the Riemann-Liouville fractional derivative of order $\alpha$ with respect to the variable $t$. The definition of this derivative is
\begin{align}\label{eq:2}
\frac{\partial^\alpha {u(t,x)}}{\partial{t}^\alpha}=\left\{
\begin{array}{lll}
  \frac{\partial^n {u}}{\partial{t}^n} & , & \alpha=n, \\
  \frac{1}{\Gamma(n-\alpha)} \frac{\partial^n}{\partial{t}^n} \int^{t}_{0}(t-s)^{n-\alpha-1}u(s,x)ds & , & 0\leq n-1<\alpha<n,
\end{array}\right.
\end{align}
here $n\in N$, $\frac{\partial^n}{\partial{t}^n}$ is the usual partial derivative of integer order $n$ with respect to $t$.\par
The nonlinear time-fractional convection-diffusion equation is obtained from the following nonlinear convection-diffusion equation by replacing the time derivative by fractional derivative.
$$u_{t}=(D(u)u_{x})_{x}+P(u)u_{x},$$
where $u=u(t,x),u_{t}=\frac{\partial{u}}{\partial{t}},u_{x}=\frac{\partial{u}}{\partial{x}}$, $D(u),P(u)$ are referred to  the diffusivity and convective terms respectively.
The nonlinear convection-diffusion equation arises in many areas of science and engineering such as being used to model the evolution of thermal waves in plasma[27].

Our paper is organized as follows. In Section 2, we introduce Lie symmetry analysis for FDEs and deduce the infinitesimals of symmetries for time-fractional convection-diffusion equation (1) by considering the different conditions for $D(u)$ and $P(u)$. In Section 3, the similarity reductions for Eq.(1) are presented by Lie symmetries obtained in section 2. Some invariant solutions of Eq.(1) are given.

\section{Symmetry analysis of Eq.(1)}

Next we use Lie symmetry analysis for Eq.(1). Let us assume that Eq.(1) is invariant under the following one parameter($\epsilon$) continuous transformations
\begin{equation}
\begin{aligned}
x^{*}&=x+\epsilon\xi(t,x,u)+o(\epsilon),\\
t^{*}&=t+\epsilon\tau(t,x,u)+o(\epsilon),\\
u^{*}&=u+\epsilon\eta(t,x,u)+o(\epsilon),\\
u^{*}_{x^{*}}&=u_x+\epsilon\eta^x+o(\epsilon),\\
u^{*}_{x^{*}x^{*}}&=u_{xx}+\epsilon\eta^{xx}+o(\epsilon),\\
\frac{\partial^{\alpha}u^*}{\partial t^{*\alpha}}&=\frac{\partial^{\alpha}u}{\partial t^{\alpha}}+\epsilon\eta^t_{\alpha}+o(\epsilon),
\end{aligned}
\end{equation}
where $\xi,\tau,\eta$ are infinitesimals and $\eta^x,\eta^{xx},\eta^t_{\alpha}$ are extended infinitesimals of orders 1,2 and $\alpha$ respectively. The explicit expressions for $\eta^x,\eta^{xx}$ are
\begin{align}
\eta^{x}&=D_{x}\eta-u_{x}(D_{x}\xi)-u_{t}(D_{x}\tau),\nonumber\\
\eta^{xx}&=D_{x}(\eta^{x})-u_{xx}(D_{x}\xi)-u_{xt}(D_{x}\tau),\nonumber
\end{align}
where symbol $D_{x}$ stands the total derivative operator with respect to $x$,
$$D_{x}=\frac{\partial}{\partial{x}}+u_{x}\frac{\partial}{\partial{u}}+u_{xx}\frac{\partial}{\partial{u_{x}}}+u_{tx}\frac{\partial}{\partial{u_{t}}}+\cdots,
$$
with infinitesimal generator $X=\xi(t,x,u)\frac{\partial}{\partial{x}}+\tau(t,x,u)\frac{\partial}{\partial{t}}+\eta(t,x,u)\frac{\partial}{\partial{u}}$. Since the lower limit of the integral in Eq.(1) is fixed and, therefore it should be invariant with respect to the transformations(3). Such invariance condition arrives at
\begin{equation}
\tau(t,x,u)|_{t=0}=0.
\end{equation}
The $\alpha$th extended infinitesimal related to Riemann-Liouville fractional time derivative reads
\begin{equation}
\eta^{t}_{\alpha}=D^{\alpha}_{t}\eta+\xi (D^{\alpha}_{t}u_{x})-D^{\alpha}_{t}(\xi u_{x})+D^{\alpha}_{t}(u D_{t}\tau)-D^{\alpha+1}_{t}(\tau u)+\tau (D^{\alpha+1}_{t}u).
\end{equation}
Here the symbol $D_{t}$ stands the total derivative operator with respect to $t$, i.e., $$D_{t}=\frac{\partial}{\partial{t}}+u_{t}\frac{\partial}{\partial{u}}+u_{xt}\frac{\partial}{\partial{u_{x}}}+u_{tt}\frac{\partial}{\partial{u_{t}}}+\cdots,$$ and the operator $D^{\alpha}_{t}$ is the total fractional derivative operator. By means of the generalized Leibnitz rule [12]
\begin{equation*}
D^{\alpha}_{t}(f(t)g(t))=\sum^{\infty}_{n=0}\binom{\alpha}{n}(D^{\alpha-n}_{t}f(t))(D^{n}_{t}g(t)),\ \ \ \alpha>0,
\end{equation*}
where
\begin{equation*}
\binom{\alpha}{n}=\frac{(-1)^{n-1}\alpha\Gamma(n-\alpha)}{\Gamma(1-\alpha)\Gamma(n+1)},
\end{equation*}
the above Eq.(5) can be written as
\begin{equation*}
\eta^{t}_{\alpha}=D^{\alpha}_{t}\eta-\alpha(D_t\tau)\frac{\partial^{\alpha}u}{\partial t^{\alpha}}-\sum^{\infty}_{n=1}\binom{\alpha}{n}(D^{n}_{t}\xi)(D^{\alpha-n}_{t}u_{x})-\sum^{\infty}_{n=1}\binom{\alpha}{n+1}(D^{n+1}_{t}\tau)(D^{\alpha-n}_{t}u).
\end{equation*}
Furthermore, using the generalized chain rule for a compound function [28]
\begin{equation*}
\frac{d^{\alpha}u(v(t))}{dt^{\alpha}}=\sum^{\infty}_{n=0}\sum^{n}_{k=0}\binom{n}{k}\frac{(-v(t))^{k}}{n!}\frac{\partial^{\alpha}(v^{n-k}(t))}{\partial{t}^{\alpha}}\frac{d^{n}u(v(t))}{dv^{n}}
\end{equation*}
along with the above generalized Leibnitz rule with $f(t)=1$,
the first term $D^{\alpha}_{t}\eta$ in $\eta^{t}_{\alpha} $ can be written as
\begin{equation*}
D^{\alpha}_{t}\eta=\frac{\partial^{\alpha}\eta}{\partial{t}^{\alpha}}+\eta_{u}\frac{\partial^{\alpha}u}{\partial{t}^{\alpha}}-u\frac{\partial^{\alpha}\eta_{u}}{\partial{t}^{\alpha}}+\sum^{\infty}_{n=1}\binom{\alpha}{n}\frac{\partial^{n}\eta_{u}}{\partial{t}^{n}}D^{\alpha-n}_{t}(u)+\mu,
\end{equation*}
where
\begin{equation*}
\mu=\sum^{\infty}_{n=2}\sum^{n}_{m=2}\sum^{m}_{k=2}\sum^{k-1}_{r=0}\binom{\alpha}{n}\binom{n}{m}\binom{k}{r}\frac{t^{n-\alpha}}{\Gamma(n+1-\alpha)}\frac{(-u)^{r}}{k!}\frac{\partial^{m}u^{k-r}}{\partial{t}^{m}}\frac{\partial^{n-m+k}\eta}{\partial{t}^{n-m+k}\partial{u}^{k}}.
\end{equation*}
Therefore
\begin{align*}
\eta^{t}_{\alpha}=&\frac{\partial^{\alpha} \eta}{\partial{t}^{\alpha}}+(\eta_{u}-\alpha
D_{t}\tau)\frac{\partial^{\alpha}u}{\partial{t}^{\alpha}}-u\frac{\partial^\alpha{\eta_{u}}}{\partial{t}^{\alpha}}+\mu+
\sum^{\infty}_{n=1}\big[\binom{\alpha}{n}\frac{\partial^{n}{\eta_{u}}}{\partial{t}^{n}}
-\binom{\alpha}{n+1}D^{n+1}_{t}(\tau)\big]D^{\alpha-n}_t u\\
&-\sum^{\infty}_{n=1}\binom{\alpha}{n}(D^{n}_{t}\xi)(D_t^{\alpha-n}u_x).
\end{align*}
\par
For the invariance of Eq.(1) under transformations(3), we have
\begin{equation}
\frac{\partial^\alpha {u^*}}{\partial{t^*}^\alpha}=(D(u^*)u^*_{x^*})_{x^*}+P(u^*)u^*_{x^*},0<\alpha<2
\end{equation}
for any solution $u=u(t,x)$ of Eq.(1). Expanding Eq.(6) about $\epsilon=0$, making use of infinitesimals and their extensions, equating the coefficients of $\epsilon$, and neglecting the terms of higher power of $\epsilon$, we obtain the following invariant equation of Eq.(1)
\begin{align}
[\eta^{t}_{\alpha}-(P'(u)u_{x}+D''(u)(u_{x})^{2}+D'(u)u_{xx})\eta
-(P(u)+2u_{x}D'(u))\eta^{x}-D(u)\eta^{xx}]|_{Eq.(1)}=0.
\end{align}
Here we assume that $D(u)$ and $P(u)$ are not equal to zero, otherwise Eq.(1) would be another equation that have been considered in [20]. Substituting the expressions for $\eta^{t}_{\alpha}$,$\eta^{x}$ and $\eta^{xx}$ into the above equation and equating various powers of derivatives of $u$ to zero, we obtain an over determined system of linear equations. They are
\begin{align}
&\xi_{t}=\xi_{u}=\tau_{x}=\tau_{u}=\eta_{uu}=0,\nonumber\\
&P(u)(\xi_{x}-\alpha\tau_{t})-P'(u)\eta-2D'(u)\eta_{x}-D(u)(2\eta_{xu}-\xi_{xx})=0,\nonumber\\
&D''(u)\eta+D'(u)(\eta_{u}-2\xi_{x}+\alpha\tau_{t})=0,\nonumber\\
&D(u)(2\xi_{x}-\alpha\tau_{t})-D'(u)\eta=0,\\
&\frac{\partial^{\alpha}{\eta}}{\partial{t}^{\alpha}}-u\frac{\partial^{\alpha}{\eta_{u}}}{\partial{t}^{\alpha}}-P(u)\eta_{x}-D(u)\eta_{xx}=0,\nonumber\\
&\binom{\alpha}{n}\frac{\partial^{n}{\eta_{u}}}{\partial{t}^{n}}-\binom{\alpha}{n+1}D^{n+1}_{t}(\tau)=0,n=1,2,\cdots\nonumber.
\end{align}

In order to solve the above system, we consider the following different conditions for $D(u)$, $P(u)$ and obtain their corresponding Lie symmetries of Eq.(1). And if $\alpha=1$, Eq.(1) becomes partial differential equation which has been considered by Oron, Rosenau[29] and Edwards[30]. Therefore, $\alpha\in (0,2)$ and $\alpha\neq1$ in our paper.\\
\textbf{Case 1 }$D(u)$ and $P(u)$ arbitrary\\
Solving the determining equations (8), we obtain the explicit form of infinitesimals
$$\xi=a_1, \tau=0, \eta=0,$$
where $a_1$ is an arbitrary constant. Hence the infinitesimal generator is
\begin{equation}\label{eq:9}
X_{1}=a_1\frac{\partial}{\partial{x}}.
\end{equation}
\textbf{Case 2 }$D(u)=u^{k}(k\neq0)$, $P(u)=\beta(\beta=\pm1)$\\
Similarly, under this assumption the following infinitesimals are obtained,
$$\xi=a_1+a_2x, \tau=a_2\frac{t}{\alpha}, \eta=a_2\frac{u}{k},$$
where $a_1$ and $a_2$ are  arbitrary constants. Hence in this case Eq.(1) admits a two-parameter group with infinitesimal generators

\begin{equation}\label{eq:10}
X_{1}=\frac{\partial}{\partial{x}},X_{2}=x\frac{\partial}{\partial{x}}+\frac{t}{\alpha}\frac{\partial}{\partial{t}}+\frac{u}{k}\frac{\partial}{\partial{u}},
\end{equation}
which are the basis of 2-dimensional Lie algebras admitted by Eq.(1). 

The other cases are listed in Table 1.

\begin{table}[h]
\newcommand{\tabincell}[2]{\begin{tabular}{@{}#1@{}}#2\end{tabular}}
 \centering
    \begin{tabular}{|l|l|l|l|l|l|l||}
      \hline
      No. & $D(u)$ &$P(u)$ & $\xi$  & $\tau$ & $\eta$  \\
      \hline
      3   & $u^{k}(k\neq0,-2,\frac{2\alpha}{1-\alpha})$   & $\beta u^{k}(\beta=\pm1)$& $a_1$ & $a_2t$& $-a_2\frac{\alpha u}{k}$ \\\hline
      4 & $u^{k}(k\neq0)$   & $\beta u^{\gamma}(\beta=\pm1,\gamma\neq k)$ & $a_1+a_2x$& $\frac{2\gamma-k}{\alpha(\gamma-k)}a_2t$ & $-\frac{a_2u}{\gamma-k}$ \\\hline
      5& $u^{-2}$  &  $\beta u^{-2}(\beta=\pm1)$& $a_1+a_3e^{-\beta x}$&$a_2t$&$\frac{a_2}{2}\alpha u+a_3\beta u e^{-\beta x}$\\\hline
      6& $u^{\frac{2\alpha}{1-\alpha}}$ & $ \beta u^{\frac{2\alpha}{1-\alpha}}(\beta=\pm1)$&$a_1$&$a_2t+a_3t^2$&$\frac{a_2(\alpha-1)}{2}u+a_3(\alpha-1)tu$ \\\hline

     7& $1$  &  $\beta(\beta=\pm1)$& $a_1$&$0$ &\tabincell{l}{$a_2u+h(t,x),$ \\ where $h(t,x)$ satisfies  \\ $\frac{\partial^{\alpha}{h(t,x)}}{\partial{t}^\alpha}=\beta h_{x}+h_{xx}$}\\\hline
     8  & $1$ & $\beta u^{\gamma}(\beta=\pm1,\gamma\neq0)$ & $a_1+a_2x$ & $\frac{2a_2}{\alpha}t$ & $-\frac{a_2}{\gamma}u$\\
    \hline
    \end{tabular}
    \caption{Infinitesimals of Eq.(1)}

\end{table}

 In the above table, $a_1,a_2,a_3$ are three arbitrary parameters. Like case 1 and case 2, the infinitesimal generators of Eq.(1) in different cases can easily obtained. Next we will use infinitesimal generators to deduce the similarity reductions and construct invariant solutions of Eq.(1). The definition of group invariant solution of  FDEs has given in [22]. Here we use it directly.
 \section{Similarity reductions of Eq.(1)}
\textbf{Case 1.}$D(u)$ arbitrary,$P(u)$ arbitrary\\
The infinitesimal generator is $X_{1}=\frac{\partial}{\partial{x}}$. The characteristic equations become
\begin{equation*}
\frac{dx}{1}=\frac{dt}{0}=\frac{du}{0},
\end{equation*}
which have two invariants $t,u.$
Thus, the similarity transformation is
\begin{equation}\label{eq:18}
u=\varphi(t).
\end{equation}
Substitution of (11) into Eq.(1) leads to $\varphi(t)$ satisfying the reduced fractional ordinary differential equation
\begin{equation}\label{eq:19}
\frac{d^\alpha {\varphi(t)}}{d{t}^\alpha}=0.
\end{equation}
Hence,the group invariant solutions of  Eq.(1) are given by
\begin{align*}
u=\left\{
\begin{array}{lll}
  c_{1}t^{\alpha-1} & , & 0<\alpha<1, \\
  c_{1}t^{\alpha-1}+ c_{2}t^{\alpha-2} & , &1<\alpha<2.
\end{array}\right.
\end{align*}
where $c_{1}$,$c_{2}$ are arbitrary constants.

\textbf{Case 2.}$D(u)=u^{k}(k\neq0,-2,\frac{2\alpha}{1-\alpha})$,$P(u)=\beta(\beta=\pm1)$ \\
The characteristic equation corresponding to $X_{2}=x\frac{\partial}{\partial{x}}+\frac{t}{\alpha}\frac{\partial}{\partial{t}}+\frac{u}{k}\frac{\partial}{\partial{u}}$ is
\begin{equation*}
\frac{dx}{x}=\frac{\alpha dt}{t}=\frac{kdu}{u}.
\end{equation*}
Solving the above equation, we get the similarity transformation
\begin{equation}
u=t^{\frac{\alpha}{k}}\varphi(\zeta), \zeta=xt^{-\alpha}.
\end{equation}
Substituting transformation(13) into Eq.(1) leads to
\begin{equation}\label{eq:22}
\frac{\partial^\alpha(t^{\frac{\alpha}{k}}\varphi(\zeta))}{\partial{t}^{\alpha}}=t^{\frac{\alpha}{k}-\alpha}(\varphi^{k}\frac{d^2\varphi}{d\zeta^2}+ k\varphi^{k-1}(\frac{d\varphi}{d\zeta})^{2}+\beta\frac{d\varphi}{d\zeta}).
\end{equation}
Because $\alpha\in(0,2)$ and $\alpha\neq1$, according to the definition of the Riemann-Liouville fractional derivative, we should consider $0<\alpha<1$ and $1<\alpha<2$ separately.

When $0<\alpha<1$,  for the similarity transformation(13) becomes
\begin{equation}\label{eq:23}
\frac{\partial^\alpha(t^{\frac{\alpha}{k}}\varphi(\zeta))}{\partial{t}^{\alpha}}=\frac{1}{\Gamma(1-\alpha)}\frac{\partial}{\partial{t}}\int^{t}_{0}(t-s)^{-\alpha}s^{\frac{\alpha}{k}}\varphi(xs^{-\alpha})ds.
\end{equation}
Let $\theta=\frac{t}{s}$, then Eq.(15) can be written as
\begin{equation*}
\begin{aligned}
\frac{\partial^\alpha(t^{\frac{\alpha}{k}}\varphi(\zeta))}{\partial{t}^{\alpha}}&=\frac{1}{\Gamma(1-\alpha)}\frac{\partial}{\partial{t}}\int^{\infty}_{1}(t-\frac{t}{\theta})^{-\alpha}(\frac{t}{\theta})^{\frac{\alpha}{k}}\varphi(\zeta \theta^{\alpha})\frac{t}{\theta^{2}}d\theta\\
&=\frac{\partial}{\partial{t}}[t^{\frac{\alpha}{k}-\alpha+1}\frac{1}{\Gamma(1-\alpha)}\int^{\infty}_{1}(\theta-1)^{-\alpha}\theta^{\alpha-\frac{\alpha}{k}-2}\varphi(\zeta\theta^{\alpha})d\theta]\\
&=\frac{\partial}{\partial{t}}[t^{\frac{\alpha}{k}-\alpha+1}(K^{1+\frac{\alpha}{k},1-\alpha}_{\frac{1}{\alpha}}\varphi)(\zeta)]\\
&=t^{\frac{\alpha}{k}-\alpha}(1+\frac{\alpha}{k}-\alpha-\alpha\zeta\frac{d}{d\zeta})[(K^{1+\frac{\alpha}{k},1-\alpha}_{\frac{1}{\alpha}}\varphi)(\zeta)]\\
&=t^{\frac{\alpha}{k}-\alpha}[(P^{1+\frac{\alpha}{k}-\alpha,\alpha}_{\frac{1}{\alpha}}(\varphi))(\zeta)],
\end{aligned}
\end{equation*}
where $P^{\tau,\alpha}_{\beta}$ is  Erdelyi-Kober fractional derivative operator and its definition is
\begin{align*}
(P^{\tau,\alpha}_{\beta}\varphi)(\zeta)=\prod^{n-1}_{j=0}(\tau+j-\frac{1}{\beta}\zeta\frac{d}{d\zeta})[K^{\tau+\alpha,n-\alpha}_{\beta}(\varphi)(\zeta)],
\zeta>0, \beta>0, \alpha>0,
n=\left\{
\begin{array}{lll}
[\alpha]+1&,&\alpha\notin N,\\
\alpha&,&\alpha\in N.
\end{array}\right.
\end{align*}
here
\begin{align*}
(K^{\tau,\alpha}_{\beta}\varphi)(\zeta)=\left\{
\begin{array}{lll}
\frac{1}{\Gamma(\alpha)}\int^{\infty}_{1}(\theta-1)^{\alpha-1}\theta^{-(\tau+\alpha)}\varphi(\zeta\theta^{\frac{1}{\beta}})d\theta&,&\alpha>0,\\
\varphi(\zeta)&,&\alpha=0.
\end{array}\right.
\end{align*}
When $1<\alpha<2$, we can also obtain $$\frac{\partial^\alpha(t^{\frac{\alpha}{k}}\varphi(\zeta))}{\partial{t}^{\alpha}}=t^{\frac{\alpha}{k}-\alpha}[(P^{1+\frac{\alpha}{k}-\alpha,\alpha}_{\frac{1}{\alpha}}(\varphi))(\zeta)],$$
by using the same method.
Then Eq.(1) can be reduced into an ordinary differential equation of fractional order
\begin{equation}\label{eq:24}
(P^{1+\frac{\alpha}{k}-\alpha,\alpha}_{\frac{1}{\alpha}}\varphi)(\zeta)=\varphi^{k}\frac{d^{2}\varphi}{d\zeta^{2}}+ k\varphi^{k-1}(\frac{d\varphi}{d\zeta})^{2}+\beta\frac{d\varphi}{d\zeta}.
\end{equation}
As for other cases, Eq.(1) can also be reduced by the similarity transformations corresponding to other infinitesimal generators. The results are as follows.

\textbf{Case 3.}  $D(u)=u^{k}(k\neq0)$,$P(u)=\beta u^{k}(\beta=\pm1)$\\
The similarity transformation $u=t^{-\frac{\alpha}{k}}\psi(\zeta)$ along with the similarity variable $\zeta=x$ reduces Eq.(1) to the nonlinear ordinary differential equation of the form

\begin{equation}\label{eq:29}
\frac{d^2\psi}{d \zeta^2}+ k\psi^{-1}(\frac{d\psi}{d\zeta})^2+\beta\frac{d\psi}{d\zeta}-\frac{\Gamma(1-\frac{\alpha}{k})}{\Gamma(1-\frac{\alpha}{k}-\alpha)}\psi^{1-k}(\zeta)=0,
\end{equation}
which is corresponding to the infinitesimal generator $t\frac{\partial}{\partial t}-\frac{\alpha u}{k}\frac{\partial}{\partial u}.$\\
\textbf{Case 4.}$D(u)=u^{k}(k\neq0)$,$P(u)=\beta u^{\gamma}(\beta=\pm1,\gamma\neq k)$

The similarity transformation $u=x^{\frac{1}{k-\gamma}}H(\omega)$ along with the similarity variable $\omega=tx^{-b},b=\frac{2\gamma-k}{\alpha(\gamma-k)}$ reduces Eq.(1) to the nonlinear ordinary differential equation of fractional order of the form
\begin{equation}
\begin{array}{r@{~}l}
&(P^{1-\alpha,\alpha}_{-1}H)(\omega)=\omega^{\alpha}[\frac{(\gamma+1)}{(k-\gamma)^{2}}H^{k+1}-b\omega(\frac{k+\gamma+2}{k-\gamma}-b)H^{k}\frac{dH}{d\omega}\\
&+b^{2}\omega^{2}(kH^{k-1}(\frac{dH}{d\omega})^{2}+H^{k}\frac{d^2H}{d\omega^2})+\frac{\beta}{k-\gamma}H^{\gamma+1}-b\beta\omega H^{\gamma}\frac{dH}{d\omega}],
\end{array}
\end{equation}
which is corresponding to the infinitesimal generator
$x\frac{\partial}{\partial{x}}+\frac{2\gamma-k}{\alpha(\gamma-k)}t\frac{\partial}{\partial{t}}+\frac{u}{k-\gamma}\frac{\partial}{\partial{u}}$.\\
\textbf{Case 5.} $D(u)= u^{-2}$,$P(u)=\beta u^{-2}(\beta=\pm1)$

The similarity transformation $u=e^{\beta x}G(\zeta)$ along with the similarity variable $\zeta=t$ reduces Eq.(1) to the nonlinear ordinary differential equation of fractional order of the form
\begin{equation}\label{eq:30}
\frac{d^\alpha G(\zeta)}{d{\zeta}^{\alpha}}=0,
\end{equation}
which is corresponding to the infinitesimal generator $e^{-\beta x}(\frac{\partial}{\partial{x}}+\beta u\frac{\partial}{\partial{u}})$.

Because the solutions of Eq.(19) is
\begin{align*}
G(\zeta)=\left\{
\begin{array}{lll}
  c_{3}\zeta^{\alpha-1} & , & 0<\alpha<1, \\
  c_{3}\zeta^{\alpha-1}+ c_{4}\zeta^{\alpha-2} & , &1<\alpha<2.
\end{array}\right.
\end{align*}
where $c_{3}$,$c_{4}$ are arbitrary constants, the group invariant solution of Eq.(1) is
\begin{align*}
u=\left\{
\begin{array}{lll}
  c_{3}t^{\alpha-1}e^{\beta x} & , & 0<\alpha<1, \\
  (c_{3}t^{\alpha-1}+ c_{4}t^{\alpha-2})e^{\beta x} & , &1<\alpha<2.
\end{array}\right.
\end{align*}\\
\textbf{Case 6.}$D(u)= u^{\frac{2\alpha}{1-\alpha}}$,$P(u)=\beta u^{\frac{2\alpha}{1-\alpha}}(\beta=\pm1)$

The similarity transformation $u=t^{\alpha-1}F(\zeta)$ along with the similarity variable $\zeta=x$ reduces Eq.(1) to the nonlinear ordinary differential equation of the form
\begin{equation}\label{eq:32}
\frac{d^2F}{d\zeta^2}+\frac{2\alpha}{1-\alpha}F^{-1}(\frac{dF}{d\zeta})^{2}+\beta \frac{dF}{d\zeta}=0.
\end{equation}
which is corresponding to the infinitesimal generator
$t^{2}\frac{\partial}{\partial{t}}+(\alpha-1)tu\frac{\partial}{\partial{u}}$.\\
\textbf{Case 7.}$D(u)=1$ and $P(u)=\beta(\beta=\pm1)$

The similarity transformation $u=e^{x}Q(\zeta)$ along with the similarity variable $\zeta=t$ reduces Eq.(1) to the nonlinear ordinary differential equation of fractional order of the form
\begin{equation}\label{eq:35}
\frac{d^\alpha Q(\zeta)}{d{\zeta}^\alpha}=(1+\beta)Q(\zeta),
\end{equation}
which is corresponding to the infinitesimal generator
$\frac{\partial}{\partial{x}}+u\frac{\partial}{\partial{u}}.$\\
Because the solution of Eq.(21) is $$Q(\zeta)=\zeta^{\alpha-1}E_{\alpha,\alpha}[(1+\beta)\zeta^{\alpha}],$$the group invariant solution of Eq.(1) is $$u=t^{\alpha-1}e^{x}E_{\alpha,\alpha}[(1+\beta)t^{\alpha}],$$\\
where $E_{\alpha,\beta}(z)$ is the Mittag-Leffler function[11].\\
\textbf{Case 8.}$D(u)=1$ and $P(u)=\beta u^{\gamma}(\beta=\pm1,\gamma\neq0)$

The similarity transformation $u=t^{-\frac{\alpha}{2\gamma}}\phi(\sigma)$ along with the similarity variable $\sigma=xt^{-\frac{\alpha}{2}}$ reduces Eq.(1) to the nonlinear ordinary differential equation of fractional order of the form
\begin{equation}\label{eq:36}
(P^{1-\frac{\alpha}{2\gamma}-\alpha,\alpha}_{\frac{2}{\alpha}}\phi)(\sigma)=\frac{d^{2}\phi}{d\sigma^{2}}+\beta\phi^{\gamma}\frac{d\phi}{d\sigma},
\end{equation}
which is corresponding to the infinitesimal generator
$x\frac{\partial}{\partial{x}}+\frac{2t}{\alpha}\frac{\partial}{\partial{t}}-\frac{u}{\gamma}\frac{\partial}{\partial{u}}$.\\

\section{Summary and discussion}
In this paper, we illustrate the application of Lie symmetry analysis to study time-fractional convection-diffusion equation. We consider the group classification of this equation for two variable functions. Eight cases are discussed. In every case Lie point symmetries are derived and
similarity reductions of this equation are performed by means of non-trivial Lie point symmetry. In some cases, the time fractional convection-diffusion equation can be transformed into a nonlinear ODE of fractional order. In other cases, this equation can be reduced to a nonlinear ODE. Some invariant solutions are given in some cases. In addition, it is easily shown that the infinitesimal generator admitted by time-fractional convection-diffusion equation in each cases can form Lie algebra and the dimension of Lie algebra is decided by the number of parameters in transformations. It is necessary to remark that case 7 is different from other cases because in that case Lie algebra is infinite dimensional and in other cases Lie algebra is finite dimensional.

Lie symmetry analysis also can be used for other time-fractional differential equations. But there are little conclusions on the symmetry property for one kind of fractional equations. This is a possible direction for future work.
\section*{Acknowledgements}

This work is supported by natural science foundation of Zhejiang Province (Grant No.Y6100611) and the national natural science foundation of China(Grant No.11371323)

\end{document}